%% file: paper.tex
\documentclass[]{fairmeta}
\usepackage{cite}
\usepackage{tabularx}
\usepackage{multirow}
\usepackage{enumitem}
\usepackage{longtable}
\usepackage{algorithm}
\usepackage{algorithmic}
\usepackage{amsmath}
\usepackage{amsfonts}
\newcolumntype{Y}{>{\centering\arraybackslash}X}

\newcommand{\R}[0]{\mathbb{R}}
\newcommand{\wpm}[2]{$#1_{\pm{#2}}$}
\newcommand{\aesmodel}[0]{\texttt{Audiobox-Aesthetics}}
\renewcommand{\cite}[1]{\citep{#1}}
\newlist{todolist}{itemize}{2}
\setlist[todolist]{label=$\square$}

\title{Meta Audiobox Aesthetics: Unified Automatic Quality Assessment for Speech, Music, and Sound}

\author[1,\dagger]{Andros Tjandra}
\author[1,\dagger]{Yi-Chiao Wu}
\author[1,\dagger]{Baishan Guo}
\author[1]{John Hoffman}
\author[1]{Brian Ellis}
\author[1]{Apoorv Vyas}
\author[1]{Bowen Shi}
\author[1]{Sanyuan Chen}
\author[1]{Matt Le}
\author[2]{Nick Zacharov}
\author[1]{Carleigh Wood}
\author[1]{Ann Lee}
\author[1]{Wei-Ning Hsu}

\affiliation[1]{FAIR at Meta}
\affiliation[2]{Reality Labs at Meta}

\contribution[\dagger]{Equal contribution}

\abstract{
    The quantification of audio aesthetics remains a complex challenge in audio processing, primarily due to its subjective nature, which is influenced by human perception and cultural context. Traditional methods often depend on human listeners for evaluation, leading to inconsistencies and high resource demands. This paper addresses the growing need for automated systems capable of predicting audio aesthetics without human intervention. Such systems are crucial for applications like data filtering, pseudo-labeling large datasets, and evaluating generative audio models, especially as these models become more sophisticated.
    In this work, we introduce a novel approach to audio aesthetic evaluation by proposing new annotation guidelines that decompose human listening perspectives into four distinct axes. We develop and train no-reference, per-item prediction models that offer a more nuanced assessment of audio quality. Our models are evaluated against human mean opinion scores (MOS) and existing methods, demonstrating comparable or superior performance. This research not only advances the field of audio aesthetics but also provides open-source models and datasets to facilitate future work and benchmarking. 
    We release our code and pre-trained model at: \url{https://github.com/facebookresearch/audiobox-aesthetics}
}

\date{January 2025}
\correspondence{Andros Tjandra at \email{androstj@meta.com}, Yi-Chiao Wu at \email{yichiaowu@meta.com}}

\metadata[Code]{\url{https://github.com/facebookresearch/audiobox-aesthetics}}

\begin{document}

\maketitle

\section{Introduction}
Measuring audio aesthetics has been a longstanding challenge in the field of audio processing. Unlike objective metrics such as frequency response or signal-to-noise ratio, audio aesthetics are inherently subjective and deeply intertwined with human perception and cultural context. This subjectivity makes it difficult to quantify audio quality in a way that is universally accepted. Traditional approaches often rely on human listeners to evaluate audio quality, which can be inconsistent and resource-intensive. As a result, there is a growing interest in developing automated aesthetic predictors. 

In the visual domain, LAION-Aesthetics project \cite{laionaes2022} trained an image aesthetic predictor and use it to curate high quality image dataset. For audio domain, an audio aesthetic predictor serves multiple purposes such as data filtering, which ensures only high-quality samples are selected for downstream tasks, and pseudo-labeling in the wild data, which efficiently annotates large datasets without manual efforts and tools for automatic evaluation. Recently, the automatic aesthetic evaluation of generative audio models has become critical because of the lack of single ground-truth reference audio in most generative applications. As these generative models become sophisticated and progress fast, there is a need for reliable, automated evaluation methods that minimize the costly human annotations during model development. An aesthetic predictor can fill this gap, offering a scalable solution for assessing the synthetic audio quality. 

There are several metrics have been utilized as a proxy or directly used as a quality assessment. In the speech domain, PESQ \cite{rix2001perceptual} and their successor POLQA \cite{beerends2013perceptual} measure utterance-level perceptual quality.  However, the ground-truth speech requirement greatly limits the applications of these intrusive methods. Many non-intrusive methods~\cite{patton2016automos, fu2018quality, lo2019mosnet, reddy2021dnsmos, kumar2023torchaudio} have been proposed to relax the restriction for the high demands of speech quality evaluation from generative tasks. To ease the out-of-domain robustness degradation caused by the limited available annotated data, most current systems focus on leveraging self-supervised learning (SSL) and system ensemble~\cite{cooper2022generalization, saeki2022utmos, qi2023ssl, baba2024utmosv2}. In the audio and music domains, Fr\'echet Audio Distance (FAD) \cite{kilgour2018fr} is widely used, where it calculates Fr\'echet distance between reference and generated audio embeddings extracted from pre-trained models, and the correlation with human perception has been shown. However, FAD is a statistic-based metric and couldn't predict utterance-level perceptual quality. The neural-based predictors~\cite{chinen2020visqol, deshmukh2024pam} trained by using the human-annotated utterance-level quality scores have been proposed to provide more fine-grained quality evaluations.

However, most of these methods focus on overall quality, which is ambiguous and barely provides insights into which parts of the audio cause the lower scores. The overall quality labels such as the mean opinion score (MOS) of the training data are also not always reliable and introduce many noises. For example, the \textit{corpus effect} and \textit{range-equalizing bias}~\cite {itu2001p1401, cooper2023investigating, huang2024mos} indicate that human perception is biased to self-preference and other samples within the test batch. Moreover, these methods are domain-specific, thus the scores cannot be calibrated across different domains. For example, although NISQA~\cite{mittag2019quality, mittag2021nisqa} and DNSMOS~\cite{reddy2021dnsmos} provide several orthogonal measurements such as noisiness and coloration in addition to the overall speech quality for more insights, these measurements are tailored to speech transmission or enhancement tasks. 

To better understand the quality of arbitrary audio, we propose a new annotation guideline with four axes named the aesthetic (AES) scores,  which cover several human perceptual dimensions. Based on the annotated AES data, we also respectively train four predictors for non-intrusive and utterance-level automatic audio aesthetic assessment (\aesmodel) models. Since the training data includes speech, music, and sounds, the proposed AES predictors are robust to arbitrary audio clips. Comprehensive data analysis and objective evaluations are conducted to show the effectiveness of the proposed AES scores and predictors compared to the standard MOS and previous works. In addition to the open-source AES predictors, we release the AES labels of a dataset named AES-Natural, collected from several public speech, sound, and music datasets to facilitate future research.

\section{Aesthetics data annotation}
\subsection{Audio Aesthetics Evaluation Axes} Existing guidelines MOS-like audio quality evaluation is usually very simple (e.g. "evaluate the quality and intelligibility of speech from 1 to 5"), which left the evaluation objective at a high-level and vague. MOS scores collected under such setting will encompass multiple quality perspectives (e.g. audio production quality, subjective quality / human preference, etc.) and could be biased against one of them depending on the raters' own understanding, leading the results to be difficult to interpret. We will demonstrate this bias later in Section \ref{tb:obj_eval}.

In this work, we propose to categorize the evaluation of audio aesthetics into different axes to reduce the ambiguity of the evaluation objective instead of using a simple MOS score. After consulting experts working in the audio production domain and conducting pilot user studies, we consolidate audio aesthetics into the following 4 axes:
  
\begin{enumerate}  
    \item  \textbf{Production Quality (PQ)} Focuses on the technical aspects of quality instead of subjective quality. Aspects including clarity \& fidelity, dynamics, frequencies and spatialization of the audio;
    \item  \textbf{Production Complexity (PC)} Focuses on the complexity of an audio scene, measured by number of audio components;
    \item  \textbf{Content Enjoyment (CE)}. Focuses on the subject quality of an audio piece. It’s a more open-ended axis, some aspects might includes emotional impact, artistic skill, artistic expression, as well as subjective experience, etc;
    \item  \textbf{Content Usefulness (CU)} Also a subjective axis, evaluating the likelihood of leveraging the audio as source material for content creation.
\end{enumerate}

\subsection{Data annotation}
Based on the above definitions, we design the annotation task to collect audio aesthetics data where we ask raters to evaluate the 4 audio aesthetics axes on a scale from 1 to 10. We provide a thorough annotation guideline (see Appendix \ref{sec:app_guidelines}) that includes detailed explanation on meaning of each axis as well as fine-grained audio quality aspects to consider when evaluating these axes. We also complement the guidelines with dozens of audio samples with score guidance to help raters internalize the expectations on what are considered high or low scores for each axes. Example annotation UI can be found in Appendix \ref{sec:app_ui}.

\textbf{Preparing audio samples.} We sample the audio data for annotation from a super-set of open-sourced and licensed datasets of 3 different audio modalities: speech, sound effects and music. Specifically, within each audio modality, we pick datasets of different quality to ensure good representation of audio samples distribution in real world so that the scores can be well generalized. We leverage information provided in metadata (e.g. speaker id, genre label, demographics, etc.) whereas possible and perform stratified sampling over these metadata columns to diversify the audio samples. In order to create a unified aesthetics score that are calibrated across different audio modalities, we shuffled audio samples of different modalities during annotation. We also apply loudness normalization\footnote{\url{https://github.com/slhck/ffmpeg-normalize}} to audio samples to remove potential confounding effect introduced by audio volume. Lastly, we collect 3 ratings per audio sample to reduce variance.

\textbf{Select high quality raters.} We design a rater qualification program to select high quality raters to work with. This is done by first curating a golden set where the scores are labeled by experts and treated as ground truth. We then invite outsourced vendor to work on the golden set and measure the correlation of between their selections and ground truth on the production quality and complexity axes. Evaluation on these two axes are more objective so we expect raters' answer to align well with ground truth. We onboard only raters with Pearson correlation $>$ 0.7 to annotation task. In the end, we are able to recruit a total number of 158 raters, representing diverse subjective opinions from the general public.
\subsection{Data visualizations} 
\begin{figure*}[ht]
    \centering
    \includegraphics[width=\linewidth]{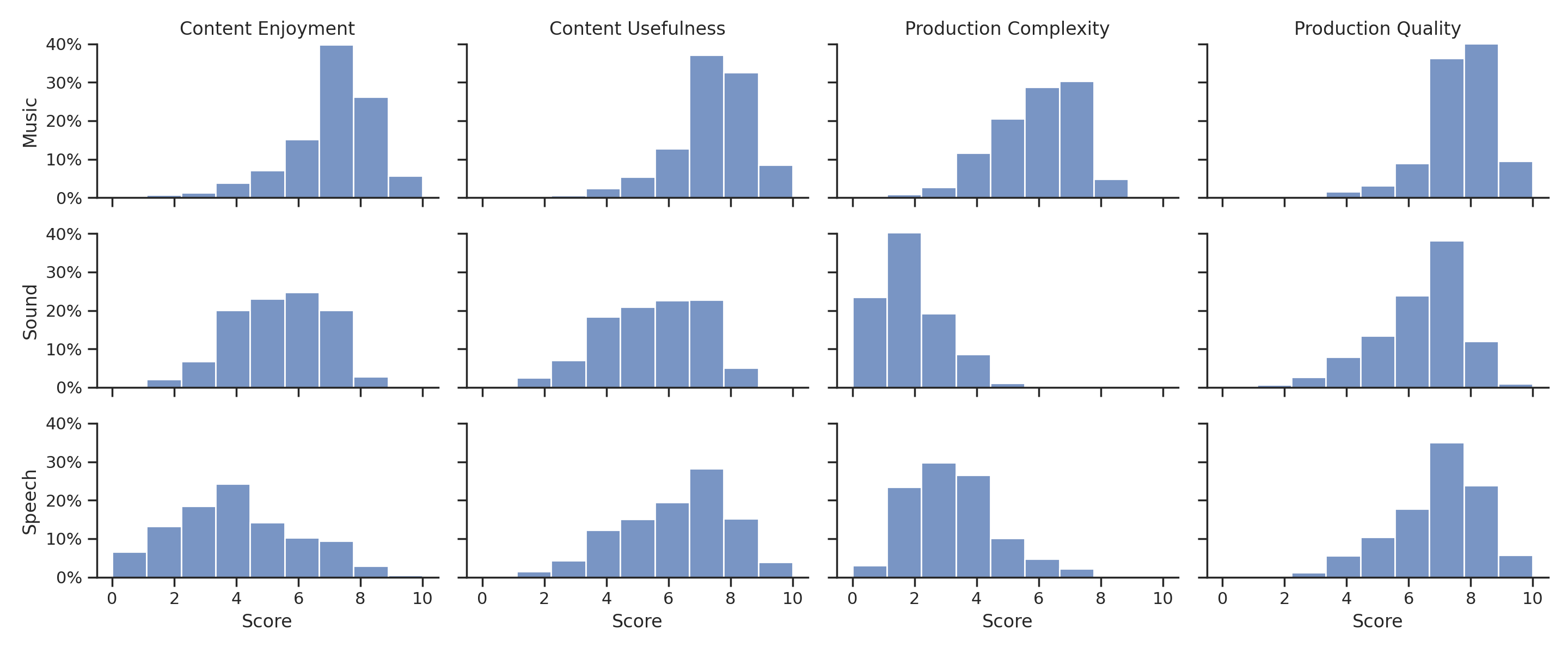}
    \caption{Aesthetic score distribution by evaluation axes and audio modalities, y-axis shows percentages in each score bucket.}
    \label{fig:score_dist}
\end{figure*}

\begin{figure}[ht]
    \centering
    \includegraphics[width=0.5\linewidth]{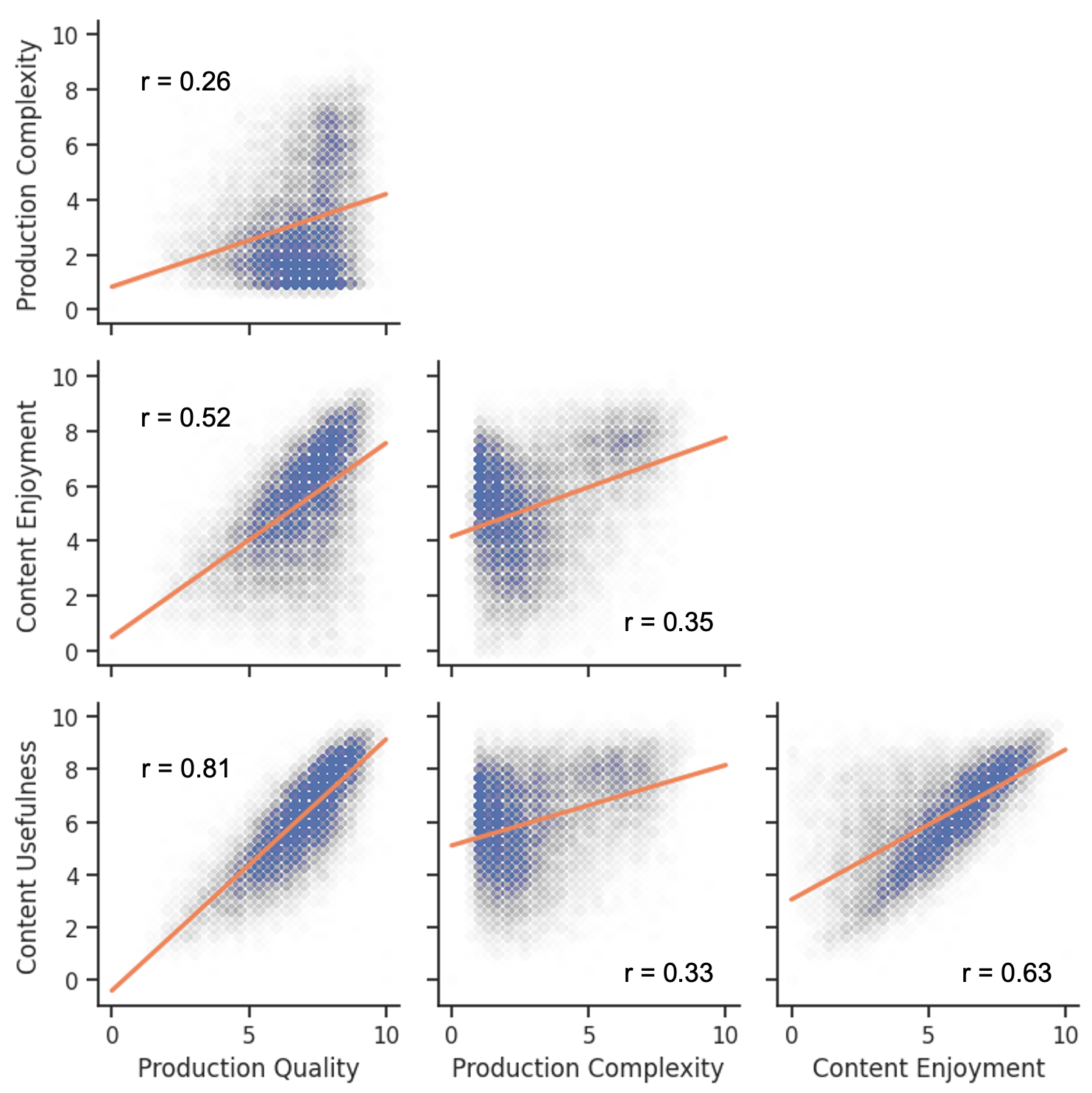}
    \caption{Correlation between different aesthetics score axes. Pearson $r$s are reported.}
    \label{fig:pairwise_plot}
\end{figure}

We annotate around 500 hours, a total number of 97k audio samples that are evenly split between speech, sound and music for training the aesthetic score predictor. Most audio samples are between 10-30 seconds. We show the score distribution by evaluation axes and audio modalities in Figure \ref{fig:score_dist}. This plot suggests for each modality we have annotated data of varying quality, while we do notice some modality-specific characteristics (e.g. music in general has higher production complexity than speech and sound effects). Figure \ref{fig:pairwise_plot} shows that scores on different axes are usually not strongly correlated, which confirms the need to de-couple aspects covered in the overall quality evaluation.

\section{Aesthetic score predictor model}

\begin{figure*}[ht]
    \centering
    \includegraphics[width=0.6\linewidth]{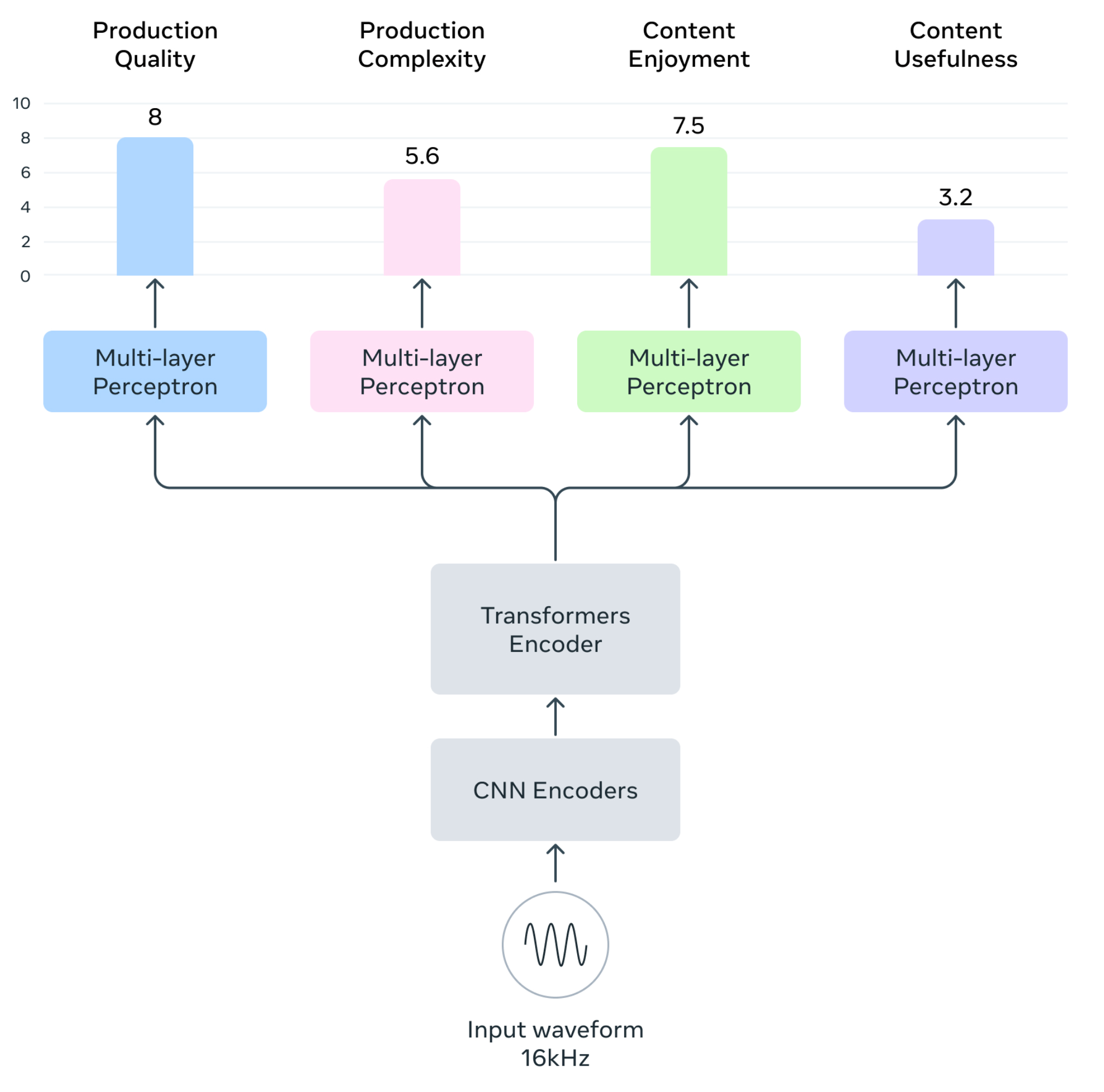}
    \caption{An overview of \aesmodel{}  architectures, input and output.}
    \label{fig:aes_model}
\end{figure*}

In this work, our aesthetic score predictor model (\aesmodel) is based on simple Transformer-based architecture. Specifically, the audio encoder is based on WavLM-based structure \cite{chen2022wavlm}, consisted of 12 Transformers \cite{NIPS2017_3f5ee243} layers with 768 hidden dimensions. Assume $h_{l,t} \in {\R}^{d}$ where $d$ is the hidden size, we extract a single embedding vector across multiple layers and timesteps to get a diverse representation from each layers. Here, we calculate audio embedding $e \in \R^{d}$ with: 
\begin{align}
    z_{l} &= \frac{w_{l}}{\sum_{l=1}^{L}{w_{l}}} \\
    \hat{e} &= \sum_{t=1}^{T}\sum_{l=1}^{L}\frac{ h_{l, t} z_{l}}{T} \\
    e &= \frac{\hat{e}}{\Vert \hat{e} \Vert_{2}}
\end{align} where $L$ is the number of Transformers layer, $T$ is the total sequence length and $w_{l} \in \R$ is the learnable scalar parameters. We project embedding $e$ through multiple multi-layer perceptron (MLP) blocks where each block consisted of multiple linear layer, layer normalization \cite{ba2016layer} and GeLU activation function \cite{hendrycks2016gaussian}. Finally, this model output predicted aesthetic scores $\hat{Y}=\{\hat{y}_{PQ},\hat{y}_{PC},\hat{y}_{CE},\hat{y}_{CU}\}$ across 4 axes. We optimize the model by minimizing the sum of mean-absolute error (MAE) and mean squared error (MSE) losses against target $Y=\{{y}_{PQ},{y}_{PC},{y}_{CE},{y}_{CU}\}$:
\begin{align}
    \mathcal{L}=\sum_{a\in\{PQ,PC,CE,CU\}}{}(y_{a}-\hat{y_{a}})^2 + |y_{a}-\hat{y_{a}}|.
\end{align} 
We show our model architecture, input and output types overview in Figure \ref{fig:aes_model}.

For data pre-processing, we re-sample all input waveform to 16kHz and single-channel. During training, we randomly sampled 10 seconds audio chunk if the audio duration is longer than 10 seconds, and we normalize output target with zero mean and one standard deviation (and apply inverse transform during inference). During inference, we predict the score using 10 seconds sliding window and calculate their weighted average, described in Algorithm \ref{alg:infer}. 
\begin{algorithm}
\caption{Audio Aesthetic Inference} \label{alg:infer}
\begin{algorithmic}[1]
\REQUIRE $x$: audio input, $sr$: sample rate
\ENSURE $y\_pred$: predicted aesthetic score

\STATE Initialize $lens \gets []$, $preds \gets []$
\STATE $stepsize \gets sr \times 10$

\FOR{$t \gets 0$ to $\text{len}(x)$ with step $stepsize$}
    \STATE $x_{now} \gets x[t \times stepsize : (t+1) \times stepsize]$
    \STATE Append $\text{AES}(x_{now})$ to $preds$
    \STATE Append $\text{len}(x_{now})$ to $lens$
\ENDFOR

\STATE $w \gets lens / \text{sum}(lens)$
\STATE $y\_pred \gets \text{sum}(preds \times w)$

\RETURN $y\_pred$
\end{algorithmic}
\end{algorithm}

\section{Objective evaluation}
\label{tb:obj_eval}
\subsection{Experimental settings}
In this section, three speech-specific and one general non-intrusive and reference-free audio quality predictors were selected as the baselines. Specifically, we adopted the P.808 MOS from DNSMOS \footnote{\url{https://github.com/microsoft/DNS-Challenge}}~\cite{reddy2021dnsmos}, PESQ from SQUIM \footnote{\url{https://pytorch.org/audio/main/tutorials/squim_tutorial.html}}~\cite{kumar2023torchaudio}, and scores from UTMOSv2 \footnote{\url{https://github.com/sarulab-speech/UTMOSv2}}~\cite{baba2024utmosv2}, which is the top system of VMC24~\cite{huang2024voicemos}, for only speech quality assessments while PAM \footnote{\url{https://github.com/soham97/PAM}}~\cite{deshmukh2024pam} was adopted for all audio types.  The selections were based on our internal preliminary evaluations. Although DNSMOS also provides the advanced P.835 score, we found that P.835 is more tailored to the speech enhancement task. We also found that SQUIM-PESQ significantly outperformed other SQUIM metrics in evaluating natural speech quality. To evaluate the effectiveness of the proposed AES scores, four models were independently trained using the corresponding AES labels. \aesmodel-PQ denotes the production-quality-predictor, \aesmodel-PC denotes the production-complexity-predictor, \aesmodel-CE denotes the content-enjoyment-predictor, and \aesmodel-CU denotes the content-usefulness-predictor.

\begin{table}[t]
\caption{Utterance- and system-level correlations between human-annotated and predicted scores on speech test set. 
}
\label{tb:mos-speech}
\centering
\selectfont
{%
\begin{tabularx}{0.9\columnwidth}{@{}p{4cm}YYYY@{}}
\toprule
\multirow{2}{*}{Model}
& \multicolumn{2}{c}{VMC22-main} & \multicolumn{2}{c}{VMC22-OOD} \\ 
& utt-PCC & sys-SRCC & utt-PCC & sys-SRCC \\ \midrule
PAM & 0.357 & 0.403 & 0.471 & 0.668 \\
DNSMOS & 0.612 & 0.773 & 0.459 & 0.615 \\
SQUIM & 0.708 & 0.711 & 0.465 & 0.515 \\
UTMOSv2 & 0.916* & 0.932* & 0.634 & 0.707 \\
\midrule
\aesmodel-PQ & 0.689 & 0.752 & 0.651 & 0.813 \\
\aesmodel-PC & -0.192 & -0.394 & -0.315 & -0.039 \\
\aesmodel-CE & \textbf{0.775} & \textbf{0.813} & \textbf{0.767} & \textbf{0.876} \\
\aesmodel-CU & 0.647 & 0.706 & 0.655 & 0.823 \\
\bottomrule
\end{tabularx}%
}
\\$^*$Data leakage: UTMOSv2 is trained using also the BVCC test set.
\end{table}

\subsection{Evaluations of public datasets}
\label{sec:evl_public}
\subsubsection{Speech}
The test sets and the MOS labels of the main and out-of-domain (OOD) tracks of VMC22~\cite{ huang2022voicemos} were adopted as the public speech datasets. The main test set includes 1066 English utterances from the BVCC~\cite{ cooper2021voices} corpus, which consists of natural, TTS, and VC speech samples from the previous Blizzard Challenges (BC), Voice Conversion Challenges (VCC), and the ESPnet-TTS~\cite{ hayashi2020espnet} published set. The OOD test set includes 540 natural and TTS Chinese utterances from the Blizzard Challenge 2019~\cite{ wu2019blizzard}. The sampling rates of both sets are 16~kHz.

As shown in Table~\ref{tb:mos-speech}, although the proposed AES models are general to arbitrary audio, the \aesmodel-PQ, \aesmodel-CE, and \aesmodel-CU predictors achieve comparable performances to other top speech-focus predictors in both utterance-level Pearson correlation coefficient (utt-PCC) and system-level Spearman's rank correlation coefficient (sys-SRCC) measurements while the general PAM predictor struggles in evaluating speech qualities. The OOD results also demonstrate the robustness of the proposed models, whose training data is mostly English but tested in Chinese. On the other hand, since the production complexity is related to how many modalities are in the samples, as expected \aesmodel-PC does not correlate to the speech qualities.

\begin{table}[t]
\caption{Utterance-level Pearson Correlation Coefficient between two types of human annotations: AES and OVL on PAM sound and music test sets.}
\label{tb:aes-mos}
\centering
\selectfont
{%
\begin{tabularx}{0.7\columnwidth}{@{}p{2cm}p{0.4cm}YYYY@{}}
\toprule
Test set & & GT-PQ & GT-PC & GT-CE & GT-CU \\ \midrule
PAM-sound & OVL & 0.496 & 0.190 & \textbf{0.581} & 0.486 \\
PAM-music & OVL & 0.778 & 0.490 & \textbf{0.848} & 0.804 \\
\bottomrule
\end{tabularx}%
}
\end{table}

\begin{table}[t]
\caption{Utterance-level Pearson Correlation Coefficient between human-annotated and predicted scores on PAM-sound.}
\label{tb:mosaes-sound}
\centering
\selectfont
{%
\begin{tabularx}{0.9\columnwidth}{@{}p{4cm}YYYYY@{}}
\toprule
Model& OVL & GT-PQ & GT-PC & GT-CE & GT-CU \\ \midrule
PAM & \textbf{0.650} & 0.408 & 0.269 & 0.542 & 0.393 \\ \midrule
\aesmodel-PQ & 0.355 & \textbf{0.617} & 0.071 & 0.406 & \textbf{0.573} \\
\aesmodel-PC & 0.092 & -0.051 & \textbf{0.654} & 0.275 & -0.098 \\
\aesmodel-CE & 0.464 & 0.318 & 0.447 & \textbf{0.638} & 0.279 \\
\aesmodel-CU & 0.396 & 0.583 & 0.058 & 0.413 & \textbf{0.573} \\
\bottomrule
\end{tabularx}%
}
\end{table}

\begin{table}[t]
\caption{Utterance-level Pearson Correlation Coefficient between human-annotated and predicted scores on PAM-music.}
\label{tb:mosaes-music}
\centering
\selectfont
{%
\begin{tabularx}{0.9\columnwidth}{@{}p{4cm}YYYYY@{}}
\toprule
Model & OVL & GT-PQ & GT-PC & GT-CE & GT-CU \\ \midrule
PAM & \textbf{0.581} & 0.568 & 0.377 & \textbf{0.699} & \textbf{0.573} \\ \midrule
\aesmodel-PQ & 0.464 & 0.587 & 0.193 & 0.449 & 0.537 \\
\aesmodel-PC & 0.251 & 0.113 & \textbf{0.710} & 0.322 & 0.096 \\
\aesmodel-CE & 0.528 & 0.487 & 0.455 & 0.661 & 0.488 \\
\aesmodel-CU & 0.465 & \textbf{0.594} & 0.221 & 0.502 & 0.558 \\
\bottomrule
\end{tabularx}%
}
\end{table}

\subsubsection{Sound and music}
The authors of PAM~\cite{deshmukh2024pam} proposed a small dataset including 500 sound (PAM-sound) and 500 music (PAM-music) samples with the corresponding text descriptions and overall quality (OVL) labels. The 500 sound/music samples consist of 100 natural audio and 400 text-to-sound/music samples from several sound/music generative models. The OVL labels were annotated by 10 raters. The sampling rates range from 16~kHz to 44.1~kHz. In addition to the OVL labels, we also hired 10 trained annotators with audio-related backgrounds to annotate this small dataset using our proposed 4 axes AES scores denoted as GT-PQ, GT-CE, GT-CU, and GT-PC for exploring the OVL and AES correlations.

Table~\ref{tb:aes-mos} shows the utt-PCC among the four AES axes and the OVL scores of the sound and music subsets. The correlations among GT-PQ, GT-CE, GT-CU, and OVL scores are high in music as expected while only GT-CE is very related to OVL in sound. It is straightforward that humans evaluate the overall audio quality according to their enjoyment level, especially for music. However, humans may have different opinions when considering the production quality or usefulness of sounds since the enjoyment of a sound is vague but the usefulness and recording quality of a sound is clear. To sum up, the results not only match our intuition on the correspondence between OVL and enjoyment but also support our argument that only overall quality is not enough to comprehensively evaluate the audio quality.

The correlation results of the human-annotated (ground-truth) and predicted scores in Table~\ref{tb:mosaes-sound} and~\ref{tb:mosaes-music} show a similar tendency as that in Table~\ref{tb:aes-mos}. Each system is most related to its training target, and the PAM-predicted scores are highly related to the enjoyment level of the sound and music. The slightly worse performance of the proposed AES predictors compared to PAM especially in music might be caused by the unseen data issue that we did not use any synthetic data to train the \aesmodel-models.

\begin{table}[t]
\caption{Utterance-level Pearson Correlation Coefficient between human-annotated and predicted scores (natural speech)}
\label{tb:aes-speech}
\centering
\selectfont
{%
\begin{tabularx}{0.8\columnwidth}{@{}p{4cm}YYYY@{}}
\toprule 
Model & GT-PQ & GT-PC & GT-CE & GT-CU \\ \midrule
PAM & 0.317 & -0.292 & 0.250 & 0.284 \\
DNSMOS & 0.662 & -0.462 & 0.598 & 0.632 \\
SQUIM & 0.660 & -0.466 & 0.570 & 0.604 \\
UTMOSv2 & 0.603 & -0.358 & 0.574 & 0.588 \\ \midrule
\aesmodel-PQ & 0.888 & -0.538 & 0.783 & 0.834 \\
\aesmodel-PC & -0.693 & \textbf{0.700} & -0.643 & -0.677 \\
\aesmodel-CE & 0.879 & -0.544 & \textbf{0.859} & \textbf{0.886} \\
\aesmodel-CU & \textbf{0.898} & -0.565 & 0.835 & 0.876 \\
\bottomrule
\end{tabularx}%
}
\end{table}

\begin{table}[t]
\caption{Utterance-level Pearson Correlation Coefficient between human-annotated and predicted scores (natural sound)}
\label{tb:aes-sound}
\centering
\selectfont
{%
\begin{tabularx}{0.8\columnwidth}{@{}p{4cm}YYYY@{}}
\toprule
Model & GT-PQ & GT-PC & GT-CE & GT-CU \\ \midrule
PAM & 0.462 & -0.022 & 0.438 & 0.443 \\  \midrule
\aesmodel-PQ & \textbf{0.728} & -0.014 & 0.552 & \textbf{0.655} \\
\aesmodel-PC & 0.106 & \textbf{0.758} & 0.297 & 0.017 \\
\aesmodel-CE & 0.492 & 0.288 & \textbf{0.763} & 0.466 \\
\aesmodel-CU & 0.676 & 0.012 & 0.571 & 0.644 \\
\bottomrule
\end{tabularx}%
}
\end{table}

\begin{table}[t]
\caption{Utterance-level Pearson Correlation Coefficient between human-annotated and predicted scores (natural music)}
\label{tb:aes-music}
\centering
\selectfont
{%
\begin{tabularx}{0.8\columnwidth}{@{}p{4cm}YYYY@{}}
\toprule
Model & GT-PQ & GT-PC & GT-CE & GT-CU \\ \midrule
PAM & 0.656 & -0.244 & 0.675 & 0.696 \\ \midrule
\aesmodel-PQ & \textbf{0.887} & -0.352 & 0.664 & 0.834 \\
\aesmodel-PC & -0.270 & \textbf{0.905} & 0.001 & -0.229 \\
\aesmodel-CE & 0.643 & -0.004 & \textbf{0.750} & 0.697 \\
\aesmodel-CU & 0.852 & -0.322 & 0.685 & \textbf{0.835} \\
\bottomrule
\end{tabularx}%
}
\end{table}

\subsection{Evaluations of AES-Natural}
Given the limited public sound and music datasets with quality labels, and most public datasets provide only the overall quality labels, we propose the AES-Natural dataset including the 4 axes labels of 2950 audio samples. Specifically, we sampled 150 utterances from EARS~\cite{richter2024ears}, 300 utterances from LibriTTS~\cite{zen19_interspeech}, and 500 utterances from Common Voice 13.0 (cv13.0)~\cite{ commonvoice:2020} to form the 950 speech subset. Since the contents of many cv13.0 files are not full-band, we resampled the cv13.0 dataset to 16~kHz. The speech subset covers different emotions, styles, accents, qualities, and sampling rates (16~kHz--48~kHz).  We also respectively sampled 451 files from MUSDB18-HQ~\cite{musdb18-hq}, 549 files from MusicCaps~\cite{agostinelli2023musiclm}, and 1k files from AudioSet~\cite{gemmeke2017audio} to form the 1k music and 1k sound subsets covering different audio types, qualities, and sampling rates (16~kHz--48~kHz). For a reasonable length for annotations, we sampled utterances mostly with 10s--30s and took the 20s--50s segments of long music. The total length of the corpus is around 11.2 hours. Each sample was annotated by 10 trained annotators with audio-related backgrounds.

According to the results of Table~\ref{tb:aes-speech} --~\ref{tb:aes-music}, we find that the production complexity is decoupled from the other three AES axes, and the \aesmodel-PC achieves a high prediction accuracy, so the proposed PC predictor provides an irreplaceable novel view for audio quality evaluations. Although production quality and enjoyment level are highly related, the proposed corresponding predictors still can distinguish production quality and enjoyment. Although the \aesmodel-CU might be replaceable by \aesmodel-CE in speech and \aesmodel-PC in sound and music, the \aesmodel-CU is still the best choice to predict usefulness given an audio-type-agnostic scenario. On the other hand, the significantly worse performance of the baselines indicates their lack of robustness to unseen data. Given the comparable performances of the AES predictors in the public datasets as shown in~\ref{sec:evl_public},  the results show the much better robustness of our \aesmodel-models.

\section{Downstream tasks}
In this section, we explore the application of an aesthetic model predictor in enhancing the performance of various downstream tasks, specifically focusing on text-to-speech (TTS), text-to-music (TTM), and text-to-audio (TTA) generation. The aesthetic model predictor is employed to refine the quality of outputs by filtering and augmenting the input data based on aesthetic scores. We conducted a series of experiments to evaluate the impact of this approach under different scenarios.
\subsection{Experimental setup}
To assess the effectiveness of the aesthetic model predictor, we designed experiments across three distinct scenarios:
\begin{enumerate}
    
\item  Baseline: We use the entire dataset without any filtering or any modification on the text prompt.
\item  Filtering strategy: We filter out part of the data where the aesthetic score is lower than $p$ percentile scores derived from aesthetic score predictions. Here, we explore threshold percentile $p = \{25, 50\}$.
\item  Prompting strategy: We incorporate aesthetic scores as text prompts to guide and control the quality of the generated outputs. For each text prompt, we simply add prefix \texttt{"Audio quality:$\hat{y}$"} where $\hat{y}=\texttt{round}(yr)/r$ and $y$ is the predicted aesthetic score from ground-truth audio and $r$ is the rounding factor. Here, we explore values of $r=\{2, 5\}$. During inference, we set the prefix as fixed string \texttt{"Audio quality: $y$"} where $y$ is percentile $p=\{50,75,90\}$ aesthetic score from training data.
\end{enumerate}
In order to obtain audio quality for filtering and prompting strategy, we predict each audio input using our \aesmodel-PQ trained model. Each scenario was tested across the three tasks: text-to-speech, text-to-music, and text-to-audio, allowing for a comprehensive evaluation of the aesthetic model predictor's utility. For every tasks, we setup a flow-matching based audio generation model similar to Audiobox-Sound \cite{vyas2023audiobox} architecture. This model controls the generation output by conditioning on text-prompt embedded from \texttt{T5-base} embedding via cross-attention mechanism. All models are trained with Adam optimizer with a learning rate $1e-4$, linearly warmup for 5k steps and linearly decayed overtime. All models are trained from scratch for 200K steps with an effective batch size of 320K frames. We use Encodec \cite{defossez2022highfi} to encode audio waveform into latent feature representation. 

\begin{table*}[ht]
\caption{This table shows objective evaluation to measures text and generated audio alignment for each downstream tasks.}
\label{tb:text-audio-obj}
\selectfont
\centering
{%
\begin{tabular}{cccccc}
\toprule 
Metric Name & & WER $\downarrow$ & CLAP-sound $\uparrow$ & CLAP-music $\uparrow$ \\ \midrule
Model & Train data (\%) & Speech & Sound & Music \\ 
\midrule
Baseline    & 100\%         & 2.95 & 0.40  & 0.36 \\
Filter $p=25$  & 75\%       & 3.37 & 0.37 & 0.36 \\
Filter $p=50$  & 50\%       & 5.06 & 0.33 & 0.36 \\
Prompt $p=50, r=2$ & 100\%  & 2.87 & \textbf{0.41} & 0.36 \\
Prompt $p=75, r=2$ & 100\%  & 2.81 & 0.40  & 0.36 \\
Prompt $p=90, r=2$ & 100\%  & 2.83 & 0.39 & 0.36 \\
Prompt $p=50, r=5$ & 100\%  & 2.84 & \textbf{0.41} & 0.36 \\
Prompt $p=75, r=5$ & 100\%  & \textbf{2.80} & \textbf{0.41} & 0.36 \\
Prompt $p=90, r=5$ & 100\%  & 2.76 & 0.40 & 0.36 \\
\bottomrule
\end{tabular}%
}
\end{table*}

\begin{table*}[ht]
\caption{This tables compares model A and model B in term of audio quality judged by human listeners. We report net win rate $[-100\%,100\%]$ and their $95\%$ confidence interval. Positive value means model A outperforms model B.}
\label{tb:audio-pair-subj}
\selectfont
\centering
{%
\begin{tabular}{cccccc}
\toprule 
Model A & Model B & Speech (\%) & Sound (\%) & Music (\%) \\ \midrule
Filter $p=25$ & Baseline & \wpm{12.93}{9.58} & \wpm{9.09}{8.92} & \wpm{9.98}{7.33} \\
Filter $p=50$ & Baseline &\wpm{15.77}{7.83} & \wpm{7.04}{9.59} & \wpm{20.99}{8.42} \\
Prompt $p=50, r=2$ & Baseline & \wpm{12.95}{9.33} & \wpm{5.38}{8.00} & \wpm{28.62}{7.96} \\
Prompt $p=75, r=2$ & Baseline & \wpm{28.44}{7.29} & \wpm{10.64}{8.75} & \wpm{46.17}{7.46} \\
Prompt $p=90, r=2$ & Baseline & \wpm{35.35}{7.59} & \wpm{15.12}{8.92} & \wpm{25.06}{7.92} \\
Prompt $p=50, r=5$ & Baseline & \wpm{10.66}{8.56} & \wpm{14.52}{9.75} & \wpm{27.66}{7.72} \\
Prompt $p=75, r=5$ & Baseline & \wpm{21.89}{8.50} & \wpm{18.40}{8.06} & \wpm{40.67}{8.75} \\
Prompt $p=90, r=5$ & Baseline & \wpm{45.07}{6.75} & \wpm{18.52}{8.92} & \wpm{23.80}{7.92} \\ \midrule
Prompt $p=75$ & Filter $p=25$ & \wpm{37.23}{8.08} & \wpm{9.12}{8.50} & \wpm{48.55}{7.83} \\
Prompt $p=75$ & Filter $p=50$ & \wpm{31.05}{8.17} & \wpm{-4.51}{9.25} & \wpm{50.19}{7.92} \\
\bottomrule
\end{tabular}%
}
\end{table*}

\subsection{Text-to-Speech}

We train our TTS model using LibriTTS \cite{zen19_interspeech} dataset. For objective evaluation, we measure transcript and generated audio truthfulness using word error rate (WER) from Whisper-V2 \cite{radford2023robust} large-v2 model in Table \ref{tb:text-audio-obj}.

\subsection{Text-to-Music}
We train our TTM model using an internally licensed dataset of high quality music collections, totalling around 18k hours. For training efficiency, we chunked the audio into 10 seconds randomly during training. During inference, we generate 10 seconds audio using 200 synthetic captions simulating our training dataset writing style with LLM-LLama3 \cite{llama2024}. For objective evaluation, we evaluate CLAP-music score using Laion-CLAP \cite{laionclap2023} to measure the alignment between text prompt and generated music in Table \ref{tb:text-audio-obj}. 

\subsection{Text-to-Audio}
We train our TTA model using mitigated version of AudioCaps \cite{kim-etal-2019-audiocaps}, totalling around 63 hours. During inference, we generate 10 seconds audio using AudioCaps test captions. For objective evaluation, we evaluate CLAP-sound score using Laion-CLAP to measure the alignment between text prompt and generated audio in Table \ref{tb:text-audio-obj}. 

\subsection{Pairwise audio quality evaluation}
To compare which strategy leads the best audio quality, we run a subjective evaluation using pairwise comparison protocols with 3 choices (+1 if model A better, -1 if model B better, 0 if similar). For each datasets, we randomly sample 200 generated audio from each corresponding test set for this evaluation. Each audio pairs are evaluated by 3 human listeners. We do bootstrap re-sampling 1,000 times  and get their percentile 2.5\% and 97.5\% result as 95\% confidence interval. The net win rate between model A vs B ranges from -100\% to 100\%. 

Table \ref{tb:audio-pair-subj} shows pairwise comparison results across all tasks with various different scenarios. In all cases, both filtering and prompting scenario always provides better audio quality compared to baseline. 

\subsection{Prompting v.s. filtering}
Based our objective and subjective evaluation across different tasks, we observed that improving audio quality generation using prompting is a better option compared to filtering. The main reason is while filtering approach keep only high quality data, but the amount of data seen by the generation model also decreased linearly compared to the quality threshold. It brings negative impact on the audio alignment objective metric such as WER or CLAP similarity scores. In term of subjective evaluation, prompting wins all versus filtering, and the prompting also keeps similar audio alignment objective compared to baseline, which means it preserved the controllability and improved the output quality simultaneously. 

\section{Conclusion}
In this paper, we presented the first approach to measuring aesthetic scores on audio modality, including speech, music, and sound domain. First, we explain how we factorize aesthetic scores in different axes to reduce ambiguity compared to standard MOS scores. Later, we train several models using our annotated dataset. We run several experiments to compare our aesthetic annotation score with several existing benchmarks and models. We also analyze the results and shows that MOS highly correlated towards audio enjoyness metric. Lastly, we set out several downstream tasks such as TTS, TTA, and TTM that utilize aesthetic score and show adding aesthetic score as a prompt improved generation quality while maintaining the alignment metric.

\clearpage
\newpage
\bibliographystyle{assets/plainnat}
\bibliography{paper}

\clearpage
\newpage
\beginappendix

\appendix
\input{sections/appendix}

\end{document}

%% file: sections/appendix.tex
\section{Aesthetics score annotation UI}
\label{sec:app_ui}
\begin{figure}[ht]
    \centering
    \includegraphics[width=\linewidth]{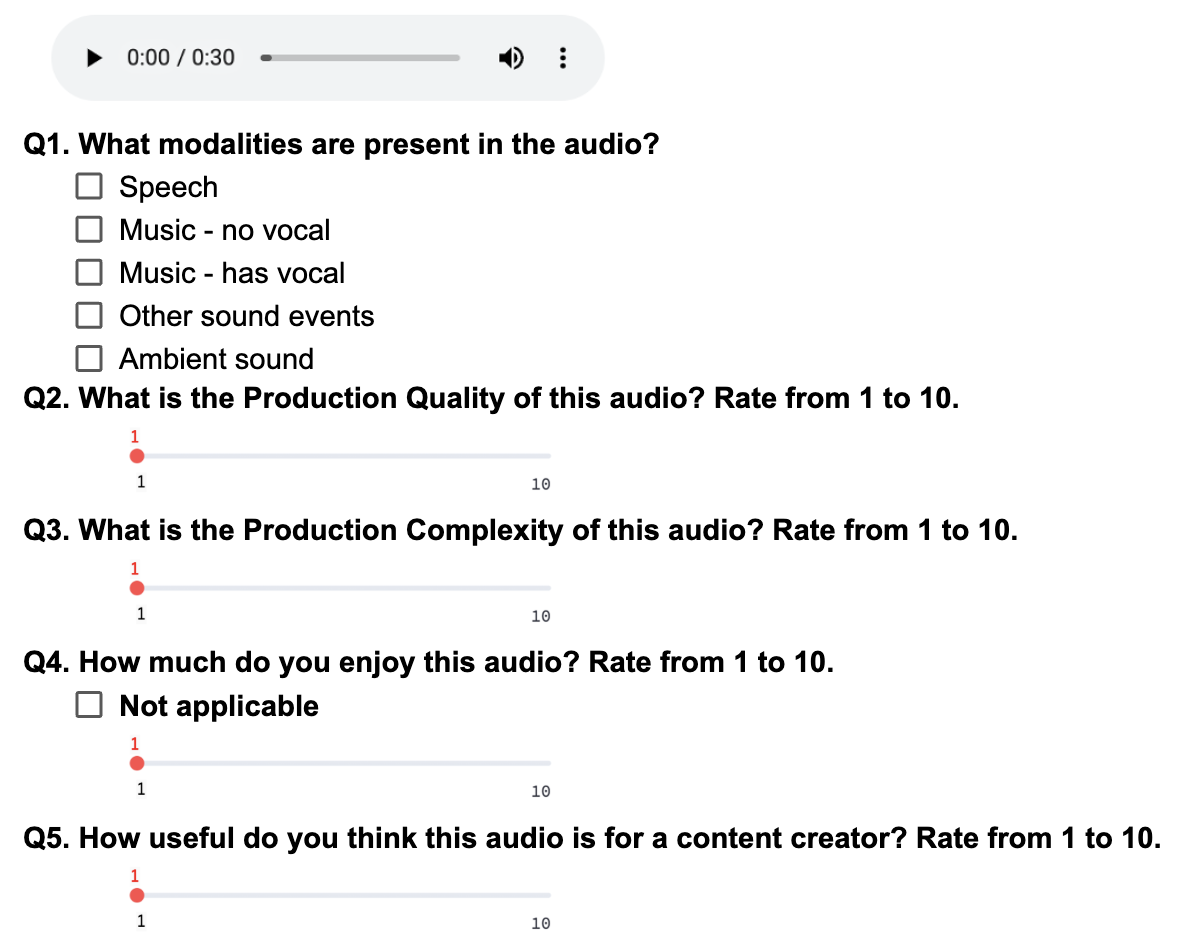}
    \caption{Aesthetic score annotation UI.}
    \label{fig:annotation_ui}
\end{figure}

\section{Aesthetics score annotation guidelines}
\label{sec:app_guidelines}

\begin{table*}[ht]
    \centering
    \caption{Audio aesthetics score annotation guidelines.}
\begin{tabular}{p{15cm}}
\toprule\toprule
\textbf{Reject the audio if}\\
\begin{enumerate}[leftmargin=.75cm]
    \item Audio does not load properly;
    \item Has violating content;
    \begin{enumerate}
        \item Hate speech - Violent, dehumanizing speech targeting a person or group of people on the basis of their protected characteristic(s), slurs;
        \item Sexual content;
        \item Other usage of strong \& explicit language (e.g., profanities, obscenities,implied threats).
    \end{enumerate}
\end{enumerate}\\

\textbf{Q1. What modalities are present in the audio?}\\
\begin{todolist}[leftmargin=.75cm]
  \item Speech;
  \item Music - no vocal;
  \item Music - has vocal;
  \item Other sound events;
  \item Ambient sound;
\end{todolist}\\

\textbf{Q2. What is the Production Quality of this audio? Rate from 1 to 10.} \\
You should focus only on \textbf{technical aspects of quality} instead of subjective quality. We want you to rate the quality based on aspects including \underline{clarity \& fidelity}, \underline{dynamics}, \underline{frequencies} and \underline{spatialization of the audio}.\\
\\
More specifically,\\
\begin{enumerate}[leftmargin=.75cm]
    \item \textbf{Clarity and Fidelity}: High quality audio should have clear, crisp sound with minimal distortion, noise, or artifacts:
    \begin{itemize}
        \item The instruments, vocals, and other elements should be well-defined and easily distinguishable / intelligible;
        \item No microphone noise / other white noise;
        \item No distortions of vocals and other elements;
        \item No other audio artifacts (e.g. hissing, buzzing, shrill, etc.)
    \end{itemize}
    \item \textbf{Dynamics}: High quality audio should maintain an appropriate dynamic range, encompassing both quiet and loud passages with clarity and impact:
    \begin{itemize}
        \item Transitions between different dynamic levels are smooth and natural, with gradual changes in volume rather than abrupt jumps;
        \item The subtle nuances and variations in volume should be well-preserved.
    \end{itemize}
    \item \textbf{Frequencies}: High quality audio should exhibit a balanced and natural frequency response across the entire spectrum, with each frequency range contributing harmoniously to the overall sound:
    \begin{itemize}
        \item Low frequency bass sound should be well-defined without muddiness or boominess;
        \item High frequency sound should be crisp and detailed without harshness or sibilance.
    \end{itemize}
    \item \textbf{Spatialization}: For multi-channel audio, the spatialization of audio elements within the stereo field should be well-defined and appropriately positioned. This creates a sense of depth and dimensionality, enhancing the listening experience.
    \item \textbf{Overall technical proficiency}: During the recording, mixing, and mastering of audio, whether it exhibits skillful application of techniques and tools to achieve high-quality sound reproduction and optimal sonic results.
\end{enumerate}\\

\bottomrule\bottomrule
\end{tabular}
\end{table*}

\begin{table*}[ht]
    \centering
    \caption{(continued) Audio aesthetics score annotation guidelines.}
\begin{tabular}{p{15cm}}
\toprule\toprule

\textbf{Q3. What is the Production Complexity of this audio?, Rate from 1 to 10.} \\
\begin{itemize}[leftmargin=.75cm]
    \item Complex production means that there are many audio components (may or may not from the same audio modality) mixed together
    \begin{itemize}
        \item e.g. You can think of a piece of podcast audio with speech, music and sound effects mixed together as high complexity. Alternatively, a piece of symphony with many instruments playing together should also be considered as complex;
    \end{itemize}
    \item Simple production means with few audio elements and components 
    \begin{itemize}
        \item e.g. Only one speaker speaking no other audio events, Piano sound only, etc.
    \end{itemize}
\end{itemize}\\

\textbf{Q4. How much do you enjoy this audio? Rate from 1 to 10.} \\
In this question we ask you to rate the subject quality, it’s an open-ended question as everyone has their own preferences and tastes. However, there are some directions / aspects that you can consider when appreciate these audio pieces: \\
\begin{enumerate}[leftmargin=.75cm]
    \item \textbf{Emotional Impact}: This means the ability of the audio to evoke emotions, convey mood, and connect with the listener. Are you able to resonate with the expressiveness / emotive quality of the audio piece?
    \item \textbf{Artistic Skill}: If the audio is for entertainment purpose (e.g. clips of music / podcast / audiobook), then the performer / speaker should demonstrate high artistic / professional skills;
    \item \textbf{Artistic Expression}: Focus on the creativity and originality in the audio. Is it innovative and gives you a unique audio experience?
    \item \textbf{Subjective Experience}: Ultimately, the subjective experience of the listener is paramount when rating the aesthetic/subjective quality of audio. Your score should reflect your personal preferences, individual taste, and emotional response.
\end{enumerate}
\\
\textbf{Q5. How useful do you think this audio is? Rate from 1 to 10.} \\
For usefulness, imagine you are a YouTube or Instagram content creator, and want to generate popular and high quality audio-visual clips (movie level quality), how likely would you be able to use this audio as source material to create some contents?\\
\bottomrule\bottomrule
\end{tabular}
\end{table*}